\documentclass[twocolumn,prl,showpacs,superscriptaddress]{revtex4}
\usepackage{graphicx}
\usepackage{subfigure}
\usepackage{amsmath}

\begin{document}

\newenvironment{itemizess}{
\begin{itemize}
  \setlength{\itemsep}{0pt}
  \setlength{\parskip}{0pt}
  \setlength{\parsep}{-2pt}
}{\end{itemize}}

\title{Quantitative analysis of several random lasers}
\author{Karen L. van der Molen and Allard P. Mosk}
\affiliation{Complex Photonic Systems, MESA$^+$ Institute for
Nanotechnolgy and Department of Science and Technology\\ University
of Twente, PO Box 217, 7500 AE Enschede, The Netherlands.}
\author{Ad Lagendijk}
\affiliation{FOM Institute for Atomic and Molecular Physics,
Kruislaan 407, 1098 SJ Amsterdam, The Netherlands}

\begin{abstract}
We prescribe the minimal set of experimental data and parameters
that should be reported for random-laser experiments and models.
This prescript allows for a quantitative comparison between
different experiments, and for a criterion whether a model predicts
the outcome of an experiment correctly. In none of more than 150
papers on random lasers that we found these requirements were
fulfilled. We have nevertheless been able to analyze a number of
published experimental results and recent experiments of our own.
Using our method we determined that the most intriguing property of
the random laser (spikes) is in fact remarkably similar for
different random lasers.
\end{abstract}

\maketitle

The research on strongly scattering media with optical gain, i.e.
random lasers, was initiated in 1968 by a pioneering paper of
Lethokov \cite{Letokhov1968}. He predicted that amplification
through stimulated emission is possible in a random medium with
gain. Since this prediction many papers on random lasers have been
published, of which we mention only a few experimental
\cite{Lawandy1994,Mujumdar2004,Anglos2004,
Cao2000,Wu2004,Frolov1999,Polson2005,Noginov1995,Milner2005} and
theoretical\cite{Burin2001,Beenakker1996,Florescu2004,Jiang2004,Patra2002,Deych2005,Angelani2006,Yamilov2005}
papers. Typical laser phenomena, like a threshold in the power
conversion and spectral narrowing, have been observed in random
lasers. In some cases, sharp features (spikes) in the emitted
spectrum occurred. The width of these spikes resembles the width of
the output of cavity lasers.

Models that have been proposed to explain spikes in the emitted
spectrum include a local cavity model with interference in a random
laser \cite{Cao2000}, also referred to as the local mode model, and
the lucky-photon model without interference taken into account
\cite{Mujumdar2004}, also referred to as the open mode model. As of
yet, no consensus exists which physical mechanisms underly spike
formation in random lasers, and it is therefore not clear which
parameters influence this formation most.\cite{Molen2006b} A
comparison between different experimental studies is very difficult,
as the experiments have many parameters not all of which are
described completely in literature.

In this Letter we propose a set of experimental data and parameters
to be reported in publications on random laser experiments. This set
of data allows for a comparison between different experiments,
between different theories, and between experiments and theory. The
set of data we suggest can be divided in sample properties and
experimental data. After we describe this set of data we will report
on an analysis of published experimental results, new experiments of
our own, and models using our prescript.

\begin{table*}[t] \footnotesize
  \centering
  \makebox[100pt]{%
\begin{tabular}{r@{}r r@{}lr@{}lrr@{}l@{}lrr@{}lr@{}lrrr@{}l}
    \hline
    no. & [ref.]  &
     \multicolumn{2}{c}{$\ell$ }&  \multicolumn{2}{c}{$\ell_a$} &
     \multicolumn{1}{c}{scatt.}& \multicolumn{3}{c}{scatt. density}
    & gain & \multicolumn{2}{c}{$\ell_g$}&\multicolumn{2}{c}{$A$} & $\lambda_p$ &
    $t_p$& \multicolumn{2}{c}{$I$}
    \\
    & &  \multicolumn{2}{c}{[$\mu$m]}&  \multicolumn{2}{c}{[$\mu$m]} &
    \multicolumn{1}{c}{}&\multicolumn{3}{c}{[m$^{-3}$]}
    & $^*$ & \multicolumn{2}{c}{[$\mu$m]} & \multicolumn{2}{c}{[mm$^2$]} & [nm] &
    [ps]& \multicolumn{2}{c}{[MW/mm$^2$]}
   \\
\hline
1 &\cite{Lawandy1994} &  200& & 172&.5 & TiO$_2$ & 2 & .8 & $\times$ 10$^{10}$ & R640P & \multicolumn{2}{c}{n.p.} &2&.5 & 532 & 7000 &0&.15 \\
2 &\cite{Mujumdar2004} &  87& .8& 18&.0 & ZnO & \multicolumn{3}{c}{n.p.} & R6G & \multicolumn{2}{c}{n.p.}&0 & .0035 & 532 & 25 & \multicolumn{2}{c}{n.p.}\\
3 &\cite{Mujumdar2004} &  538& & 18&.0 &  ZnO & \multicolumn{3}{c}{n.p.} &R6G & \multicolumn{2}{c}{n.p.}&0 & .0035 & 532 & 25 &\multicolumn{2}{c}{n.p.}\\
4 & \cite{Anglos2004} &  9& .5& \multicolumn{2}{c}{n.a.} & ZnO & 6 & .55 & $\times$ 10$^{19}$ & ZnO  &\multicolumn{2}{c}{n.p.}& 5 & & 248 & 5& 13&.4 \\
5& \cite{Cao2000} &  8& .5& 86&.3 &  ZnO & 2 & .5 & $\times$ 10$^{11}$ & R640P & \multicolumn{2}{c}{n.p.}&\multicolumn{2}{c}{n.p.$^\flat$} & 532 & 25& 36000&\\
6 &\cite{Cao2000} &  3&.0 & 86&.3 &  ZnO & 1 &  & $\times$ 10$^{12}$ & R640P  & \multicolumn{2}{c}{n.p.}& \multicolumn{2}{c}{n.p.$^\flat$} & 532 & 25& 20714&\\
7  &\cite{Cao2000} &  4& .9& 86&.3 & ZnO & 6 &  & $\times$ 10$^{11}$ & R640P  & \multicolumn{2}{c}{n.p.}&\multicolumn{2}{c}{n.p.$^\flat$} & 532 & 25&24286& \\
8  &\cite{Wu2004} &  2& & n.p.&  & ZnO & 3 & .66 & $\times$ 10$^{18}$ & ZnO  &\multicolumn{2}{c}{n.p.}& 0 &.00005 & 355 & 20&11&\\
9&  \cite{Frolov1999} & $\geq$ 500& & 89&.8 &  SiO$_2$ & 5 & .23 & $\times$ 10$^{19}$ & R6G  &\multicolumn{2}{c}{n.p.}& \multicolumn{2}{c}{var.} & 532 & var. &0&.1\\
10& \cite{Frolov1999} &  $\geq$ 500& & 89&.8 &  SiO$_2$ & 5 & .23 & $\times$ 10$^{19}$& R6G  &\multicolumn{2}{c}{n.p.}& \multicolumn{2}{c}{var.} & 532 & var.&0&.15\\
11 &\cite{Polson2005} &   12& & 89&.8 &  TiO$_2$ & 8 & .6 & $\times$ 10$^{9}$& R6G  &\multicolumn{2}{c}{n.p.}& \multicolumn{2}{c}{var.} & 532 & 100&\multicolumn{2}{c}{n.p.}\\
12 &\cite{Polson2005} &   12& & 89&.8 & TiO$_2$  & 8 & .6 & $\times$ 10$^{9}$& R6G  &\multicolumn{2}{c}{n.p.}& \multicolumn{2}{c}{var.} & 532 & 100&\multicolumn{2}{c}{n.p.}\\
13 &\cite{Noginov1995} &  n.p.& & 15& & Al$_2$O$_3$ &\multicolumn{3}{c}{n.p.}& R6G  &\multicolumn{2}{c}{n.p.}& \multicolumn{2}{c}{n.p.} & 532 & 10000& \multicolumn{2}{c}{n.p.}\\
14 &\cite{Molen2006b} & 0 & .6 & 22 && GaP & & & $^\sharp$& R640P &
12 &&0
&.000003 & 567 & 3000 & 0&.016 \\
 \hline

\end{tabular}
}

\footnotesize n.p. : not presented, var.: different values were
listed, introducing ambiguity about what value is relevant\\ $^*$
R640P = Rhodamine 640 perchlorate, and R6G = Rhodamine 6G. All the
dyes are dissolved in methanol, except for numbers 9 and 10, here
ethylene glycol is used. $^\sharp$ Porosity GaP 45\% air
 $^\flat$ Based on the used lens and
pump wavelength we estimate $A = 5.6 \times 10^{-6}$ mm$^2$

  \caption{The sample properties and experimental details of several random lasers from literature.
   Listed are the transport mean free path $\ell$, absorption length $\ell_a$, the scatter material and
   the density of the scatterers, the gain material and the gain length $\ell_g$, the size of the focus of
   the pump light on the sample $A$, the pump wavelength $\lambda_p$, the pulse duration of the pump
   pulse $t_p$ and the pump fluence of the spectrum under consideration $I$.}\label{tab,1}
\end{table*}

At least the following optical and material properties of the sample
are needed for a comparison: 
\begin{itemizess}
  \item[-] transport mean free path $\ell$ (including the measurement method), as it provides key information about the strength of scattering
  \item[-] absorption length of the pump light $\ell_a$, as it provides information about how far the pump light can travel inside the random laser
  \item[-] characterization of the scatterers (material, density, and thickness of the sample), for information about, e.g., damage threshold and heat conductivity
  \item[-] gain material (material, and minimum gain length)
  \item[-] presence (absence) of window or substrate surrounding the sample
\end{itemizess}

At least the following experimental details are required:
\begin{itemizess}
  \item[-] focus area $A$ of the pump beam on the sample, as it provides information of the size and
  shape (together with $\ell$ and $\ell_a$) of the amplified volume
  \item[-] wavelength of the pump laser $\lambda_p$
  \item[-] duration of the pump pulse $t_p$, as studies have shown that pulse duration is an important
  parameter\cite{Anglos2004,Molen2006a}
  \item[-] repetition rate of the pump laser
  \item[-] pump fluence $I$ for every published spectrum
  \item[-] integration time for every published spectrum
\end{itemizess}

Before we list the required experimental data, we briefly elaborate
on two key criteria: the occurrence of spikes and gain narrowing.
The occurrence of spikes in an emission spectrum of a random laser
is a central issue. To determine if an emission spectrum contains
spikes we take the pump fluence at a peak height (A in
Fig.~\ref{fig,analysis_spike}) and at the highest shoulder of this
peak (B in Fig.~\ref{fig,analysis_spike}). If the difference is more
than 5\% of the highest shoulder value, we count a spike. Smaller
features cannot be resolved reliably in many experiments. The width
of the spike is derived from a Lorentzian fit to the data. We
analyze each emission spectrum, count the number of spikes, and
determine the height and width of each spike. From these heights and
widths we calculate their mean value and standard deviation.
\begin{table*}[t] \footnotesize
  \centering

\begin{tabular}{rrrr@{}lr@{}lr@{}l r@{}lr@{}l r@{}lr r}
    \hline
    no. & [{\rm{ref}}]  & spikes  &  \multicolumn{2}{c}{NF} & \multicolumn{2}{c}{$I_{th}$}&  \multicolumn{2}{c}{$w$
    }&\multicolumn{2}{c}{$\sigma$($w$)
    }&
    \multicolumn{2}{c}{$h$} & \multicolumn{2}{c}{$\sigma$($h$)$^\sharp$} & $\lambda_e$ & Q \\
    &  & &  &  & \multicolumn{2}{c}{[MW/mm$^2$]} &
    \multicolumn{2}{c}{[cm$^{-1}$]}&\multicolumn{2}{c}{[cm$^{-1}$]}&
    \multicolumn{2}{c}{[\%]} & \multicolumn{2}{c}{[\%]} & [nm] & \\
      \hline
   2 &\cite{Mujumdar2004} & 13 & \multicolumn{2}{c}{n.p.} & \multicolumn{2}{c}{n.p.} & 13&.7 &
   8
    &.3&27&&24& & 585  & 1250\\
   3 &\cite{Mujumdar2004} & 11 & \multicolumn{2}{c}{n.p.} & \multicolumn{2}{c}{n.p.} &  9&.6 & 3
    &.5&28&&28& & 585 & 1780\\
    6& \cite{Cao2000} & 12 & 3 & & 7143 & & 12 &.2 &7 &.2 &30 & &16 & & 608 & 1350 \\
   7  &\cite{Cao2000} & 8 &  3 &.4 & 11429 & &  18 &.4 &10 &.7 &18 & &6 & & 608 & 894\\
   8 &\cite{Wu2004} & 10 & 3&.4&9&& 50 &.7 &20 &.2 &37 & &21 & & 375 & 526\\
   9 &\cite{Frolov1999} & 3 & 4 &.4 & 0 &.01 & 17 &.8 &17 &.8 &122&& 114& & 565 & 994\\
   10 &\cite{Frolov1999} & 13 & 4 &.4 & 0 & .01&  10 &.1 & 1 &.6 & 107&  & 159 &  & 565 & 1750\\
   11 &\cite{Polson2005} & 12 & 10 & & \multicolumn{2}{c}{n.p.} & 15 &.2 &10
    &.4 &46 & &29 & & 562 &1170\\
   12 &\cite{Polson2005} & 13 & 10& & \multicolumn{2}{c}{n.p.} &  10 & .2& 3
    &.7 &35 & &30 & & 562 & 1740 \\
    14 &\cite{Molen2006b} & 9 & 13&.3& 0&.008 &3&.0&
    1&.4&102&&170&&607 & 5490\\
    \hline
    \end{tabular}

\footnotesize n.p. : not presented.

  \caption{\label{tab,2} The experimental results for different published
   experimental emission spectra of random lasers that feature spikes, and our own emission spectrum. Listed are the number of spikes, the narrowing factor of the
   spectrum NF, the threshold of the pump fluence $I_{th}$, the mean value of the width $w$ and the standard
   deviation of the distribution of spikes, the mean value of the relative height $h$ and the standard deviation of the height distribution
   of spikes, the central wavelength of the emission spectrum under consideration $\lambda_e$, and the Q factor.}
\end{table*}
Gain narrowing can be quantified by the narrowing factor NF, defined
as the spectral width of the emitted light far above threshold
divided by the spectral width far below threshold.

In conclusion, for a thorough quantitative analysis at least the
following experimental data of the random laser are needed:
\begin{itemizess}
   \item[-] number of spikes
   \item[-] fixed spectral position of spikes
   \item[-] average width $w$ and standard deviation of width distribution (preferably in units of
   energy) of spikes
   \item[-] average relative height $h$  and standard deviation of height
   distribution of spikes
   \item[-] emission spectrum around the central emission
   wavelength $\lambda_e$.
   \item[-] narrowing factor NF
   \item[-] pump fluence at threshold, $I_{th}$
\end{itemizess}

With this quantitative framework in mind, we have done a specific
literature search to compare a number of different published
experimental results. Out of the more than 150 publications on
random lasers, we found only 8 that were complete enough for this
analysis, resulting in 13 spectra (number 1-13). The properties of
each sample and the corresponding experimental details were taken
and, where possible, translated to the properties described above.
We analyzed experimental results by scanning the published emission
spectra of the several random lasers. In addition, we analyzed our
own experimental data on a gallium phosphide random
laser\cite{Molen2006b}, the analyzed single-shot spectrum (number
14) is shown in Fig.~\ref{fig,analysis_spike}.

\begin{figure} [b]
\begin{center}
\includegraphics[width=2in]{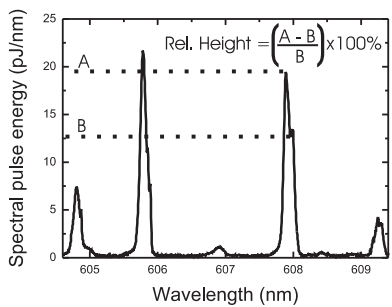}
\caption{\label{fig,analysis_spike}A measured emission spectrum of a
gallium phosphide random laser. Our method to determine whether or
not a sharp feature is a spike is displayed. We take the spectral
pulse energy A at a peak height and the spectral pulse energy B at
the highest shoulder of this peak. The formula for the relative
height is displayed in the figure. If the relative height is more
than 5\%, the feature is counted as a spike.}
\end{center}
\end{figure}

The sample properties and experimental details of our compilation
are listed in table~\ref{tab,1}. The experimental data is presented
in table~\ref{tab,2} for experiments where spikes occurred in the
emission spectrum. When we compare the different experimental data
in table~\ref{tab,2}, we notice that the average Q factor of the
laser modes (defined as $\lambda_e / w$, with $\lambda_e$ in units
of energy) is very much alike. Only our own measurement on a porous
gallium phosphide random laser (no.~14) has an average Q-factor that
is a factor 3 larger. When we compare the heights, we observe that
all the height distributions are similar, except for number 9, 10
and 14. The spectra 9 and 10 are from a very special random laser: a
photonic crystal with disorder. We conclude that, surprisingly, all
experimental results are similar within the uncertainty, except for
our own. The reason for this difference could be the very low mean
free path $\ell$ of our sample.

\begin{table}[b] \footnotesize
  \centering
\begin{tabular}{rr@{}lrr@{}l @{ $\pm$ } r@{}lr@{}l @{ $\pm$ } r@{}l}
    \hline
     & \multicolumn{2}{c}{$\ell$ } & spikes &
     \multicolumn{2}{c}{$w$}&\multicolumn{2}{c}{$\sigma(w)$}&
    \multicolumn{2}{c}{$h$} & \multicolumn{2}{c}{$\sigma(h)$}\\
    & \multicolumn{2}{c}{[$\lambda_e$]} & &
    \multicolumn{2}{c}{[cm$^{-1}$]} & \multicolumn{2}{c}{[cm$^{-1}$]}
    & \multicolumn{2}{c}{[\%]} & \multicolumn{2}{c}{[\%]} \\
    \hline
    Exp.$^*$ & 150&.1 & 13 &  13&.7 & 8
    &.3 &27&.3&23&.9 \\
    Mod.$^\natural$ & 150&.1 & 11 &  13&.9 &5
    &.4&176&&380& \\
    Exp.$^*$ & 920& & 11 &  9&.62 & 3
    &.49&27&.6&28&.3 \\
    Mod.$^\natural$ & 920& &  6 & 11 &.0 &6
    &.9&199&&317& \\
    \hline
        \end{tabular}
\\
\footnotesize $^*$ Exp. = Experiment,  $^\natural$ Mod. = Model.
  \caption{\label{tab,3}Characterization of the spikes from the experiments and the open-mode model by Mujumdar \textit{et al.} \cite{Mujumdar2004} .
   The width of the spikes predicted by the model is within one standard deviation of the
   experimental width. The prediction of the height distribution differs substantially from their experimental result.}
\end{table}

Now we will proceed to analyze the models together with the
experimental data the apply to. We only found one paper by Mujumdar
\textit{et al.} \cite{Mujumdar2004}, that showed both the outcome of
experiment and a model (the open-mode model). We determined the
characteristics of the spikes in both the experimental and
theoretical published spectra. The result is listed in
table~\ref{tab,3}. Our comparison between their model and their
experiments shows that the width distribution of their experimental
spikes extracted by us is predicted correctly by their model. The
width is not discussed explicitly in their paper. However, the
height distribution extracted by our analysis of their model differs
substantially from their experimental result.

In conclusion we have prescribed in this Letter the sets of data
needed for a thorough quantitative analysis for both random-laser
experiments and models. With these sets a comparison is possible
between experiments, and between experiments and models.
Surprisingly, all experimental results are similar within the
experimental uncertainty except for our own porous gallium phosphide
random laser.

This work is part of the research program of the 'Stichting voor
Fundamenteel Onderzoek der Materie' (FOM), which is financially
supported by the 'Nederlandse Organisatie voor Wetenschappelijk
Onderzoek' (NWO). K. van der Molen's email-address is
k.l.vandermolen@utwente.nl.

\end{document}